\documentclass[aps,pre,reprint,unsortedaddress]{revtex4-1}
\usepackage [utf8]{inputenc}
\usepackage[T1]{fontenc}
\usepackage{lmodern}
\usepackage{amsmath}
\usepackage{siunitx}
\DeclareSIUnit{\calorie}{cal}
\usepackage{hyperref}
\usepackage{mathrsfs}
\usepackage{multirow}
\usepackage{texdraw}
\usepackage{color}
\usepackage{epstopdf}
\usepackage[font=small,labelfont=bf]{caption}

\usepackage{graphicx}

\begin{document}
\thispagestyle{empty}

\title{Coarse-grained Models of Aqueous Solutions of Polyelectrolytes: Significance of Explicit Charges}

\author{Pauline Bacle}

\author{Marie Jardat}
\email{marie.jardat@sorbonne-universite.fr}

\author{Virginie Marry}

\author{Guillaume M\'eriguet}
\affiliation{Sorbonne Universit\'e, CNRS, Physico-chimie des \'electrolytes et nano-syst\`emes interfaciaux, PHENIX, F-75005 Paris, France}

\author{Guillaume Bat\^ot}
\affiliation{IFP  \'Energies Nouvelles, avenue de Bois Pr\'eau, 92852 Rueil-Malmaison Cedex, France}

\author{Vincent Dahirel}
\affiliation{Sorbonne Universit\'e, CNRS, Physico-chimie des \'electrolytes et nano-syst\`emes interfaciaux, PHENIX, F-75005 Paris, France}


\begin{abstract} 
The structure of polyelectrolytes is highly sensitive to small changes in the interactions between its monomers. In particular, interactions mediated by counterions play a significant role, and are affected by both specific molecular effects and generic concentration effects. The ability of coarse-grained models to reproduce the structural properties of an atomic model is thus a challenging task. Our present study compares the ability of different kinds of coarse-grained models: (i)~to reproduce the structure of an atomistic model of a polyelectrolyte (the sodium polyacrylate), (ii)~to reproduce the variations of this structure with the number of monomers and with the concentration of the different species. We show that the adequate scalings of the gyration radius of the polymer~$R_{\rm g}$ with the number of monomers~$N$ and with the box size~$L_{\rm box}$ are only obtained, first, if the monomer charges and the counterions are explicitely described, and second, if an attractive Lennard-Jones contribution is added to the interaction between distant monomers. Also, we show that implicit ion models are relevant only to the high electrostatic screening regime.
\end{abstract}  

\maketitle

\section{Introduction}

Charged polymers in solutions (polyelectrolytes) are ubiquitous in soft materials. For instance, they are used in paints and in food products as shear thinning and gelifying agents respectively. They can be found in soils, as a product of the biodegradation of organic matter \cite{Tochiyama_RA_2004}. They can be used in nano-technologies (nanomaterials coated with polyelectrolyte brushes \cite{Dobrynin_COCIS_2008}, electrochemical devices, solar cells \cite{Chong2011}). 

An important part of polyelectrolyte studies consists in identifying the essential factors that control the polyelectrolyte dynamics, in solution or in complex environments, and to assess their relative importance by comparing experimental results with model predictions\cite{YethirajJPCB2009,Muthukumar2017}. The models of charged polymers either consider explicitly electrostatics, or deals with polyelectrolytes as neutral polymers, as it is commonly done for instance for chromosomes\cite{care2015}. 

The capabilities of existing theories to explain and predict the behavior of a given real polyelectrolyte of interest are nevertheless limited. For instance, recent studies have revealed underestimated specific effects that should be considered even for highly diluted chains in solution\cite{Bohme-etal_2011,Salis2014,Malikova15}. Specific effects depend on the fine details of the atomic interactions within the macromolecule structure and with the surrounding species, solvent and ions.  Ion specific effects are strongly affecting the properties of polyelectrolyte gels or hydrogels\cite{Arends13}. The presence of ion-specific effects proves that purely electrostatic theories (such as Manning condensation,\cite{Manning_JCS_1969,Odijk_JPSPP_1977} or scaling theories for polyelectrolyte solutions\cite{Joanny-Barrat_1996,Dobrynin-etal_M_1995,Dobrynin-Rubinstein_2005}) omit essential interactions influencing polyelectrolyte systems. To include these effects in a polyelectrolyte model, one has to resort to modeling at a finer scale than that of the monomer, for instance using atomistic molecular dynamics.

A second class of effects that can have a strong impact comes from the difference between the properties of infinitely long polymers (the thermodynamic limit of polymer theories) and those of finite-size polymers (finite-size effects). For instance, a qualitatively different scaling behavior is observed for polymers undergoing a coil-globule phase transition, due to finite-size effects\cite{Imbert1996,Imbert1997,YethirajJCP99,Care2014,care2015}. The ability of a simple model to reproduce quantitatively the finite-size effects of the real system is an important challenge. When one determines a model using a unique length of the polymer chain, the transferability of the same potential to other polymer sizes is a crucial test to perform.

To render faithfully the dynamic properties of polyelectrolytes in solution over a long time, the use of full atom molecular dynamics is prohibitive due to the tremendous computing time needed. A common alternative is to determine a minimal mesoscopic (coarse-grained) model of a polyelectrolyte. The coarse-grained model of a polymer typically consists of superatoms which represent a group of atoms. Relatively simple models of polyelectrolytes in solution\cite{stevens95,WinklerPRL98,Chang2002} have generated a rough picture of their properties, but a more rigorous coarse-grained description where some degrees of freedom of the system are properly averaged out is more faithful\cite{Reith_JCC_2003,Vettorel2010,dAdamo_SM_2012}. Effective force fields between these superatoms, and between superatoms and simplified models of solvent molecules can be calibrated against simulations at the atomic scale\cite{An-macromolecules2019}, such that the coarse-grained model allows one to account for molecular features of the system.  Moreover, an implicit solvent description\cite{mcmillan} also allows to decrease strongly the simulation time. As the main effect of water on a polyelectrolyte chain is to screen electrostatic interactions between charged monomers and between monomers and salt ions, water can be replaced by a continuum with dielectric properties. This constitutes for example the standard description of simple electrolytes in water, proved to be reliable to describe equilibrium properties like the osmotic coefficients\cite{Dufreche02,Molina11}, or transport properties such as the electrical conductivity\cite{JardatJCP99}.  All other effects of the molecular solvent on the polyelectrolyte configuration should be included in the interactions between superatoms of the polyelectrolyte at the coarse-grained level.  

Our main goal in this paper is to compare the performance of two classes of CG models of charged polymers in implicit solvent: (i)~a model that remains a polyelectrolyte, where charge-charge interactions are explicitely accounted for, and therefore includes counterions and added salt as independent particles, and (ii)~a polymer model with fully implicit electrostatics, where ions are included into the monomer superatom. In the first case, we also investigate whether tuning the counterion-monomer interaction is enough to generate the right structural correlations between monomers, or if additional monomer-monomer interactions should be added to reproduce the polymer folding mode.   

Designing a coarse-grained model in order to predict thermodynamic, structural or dynamical properties of a system raises many questions. First, one may question the choice of the data used to calibrate the model, and the method to generate these data. In the present paper, the data come from atomistic trajectories generated by molecular dynamics. The choice of polymer isomeric configurations (in case of chiral monomers), polymer length, box size, simulation time are discussed. Second, there is an infinity of possible definitions of the superatom.  Several groups have developed algorithmic procedures to fit the parameters of coarse-grained models on the data from molecular simulations at the atomic scale. The fitting procedure can be done with a simplex method\cite{Meyer_JCP_2000} or using the iterative Boltzmann inversion\cite{Lyubartse_FD_2010,Bayramoglu12}. So far, few studies have dealt with polyelectrolytes and  derived  coarse-grained models with explicit counterions\cite{ParkJPCB2012,LiMacromolecules2012}. In the present study, the inference of the parameters of the CG models is based on the reproduction of pair distribution functions of an atomistic model. 

In order to include molecular specific effects and finite size effects of a real polymer system, we chose to study the sodium salt of the polyacrylic acid, denoted by NaPA in what follows, with a small to moderate number of monomers, 3 to 300.
This polyelectrolyte, of general formula [-CH$_2$--CH(COOH)-]$_n$, is a well-defined hydrophilic electrolyte whose charge depends on pH\cite{Laguecir2006}, that is widely used in industrial products for ion complexation\cite{Montavon_RA_2002,Kirishima_RA_2002}. Here, we focus on the case of the fully deprotonated polyacrylate chain, denoted by PA in what follows, that exists in aqueous solution whose pH is larger than 9\cite{Laguecir2006,Dolce2017}.
The atomic structure of PA monomers is chiral, and therefore the polymers we aim at modelling 
are atactic chains of randomly distributed stereoisomers.
For infinitely long polymers, the properties of the polymer may become independent of the exact distribution of stereoisomers, but for small chains, the exact choice of the configuration of each monomer shall give rise to stereospecific finite size effects. 

Coarse-grained models of the NaPA were already proposed by Reith {\em et al.} \cite{Reith_JCP_2002}, where each superatom replaced a monomer and its sodium counterion, hence each superatom was uncharged. The coarse-grained force field was fitted in order to account for pair distribution functions obtained with an atomistic molecular dynamics. These CG models allowed the authors to account faithfully for the scaling of the gyration radius~$R_{\rm g}$ with the number of monomers~$N$, in the long chain range and in the presence of an added salt,  compared to experimental determinations. However, such models of a polyelectrolyte are unable to account for transport properties depending on the charge or the amount of salt like the electrophoretic mobility of the chain or the electrical conductivity of the solution. Moreover, the study of Reith relied on relatively short atomistic trajectories, due to the limitation of computing time at that time. Our approach tackles these limitations, and challenges the use of implicit charge models of polymers for short polyelectrolytes. In order to define a relevant minimal model of NaPA, we study coarse-grained models of increasing complexity, with several definitions of superatoms. Among these models, we propose the first coarse-grained model of the sodium polyacrylate chain with explicit counterions from numerical simulations at the atomic scale. We show that attractive Lennard-Jones interactions between monomers must be accounted for to describe faithfully the behavior of the salt-free system at the atomic scale. 

Our paper is organized as follows. In Section~\ref{method}, we present the force fields and the simulation methods both at the atomic and at the coarse-grained scales. We also detail the methodology used to infer the parameters of the coarse-grained models, and the structural quantities that are computed from simulations. Then, Section~\ref{results} gives the results obtained at the atomic and coarse-grained scales, and contains discussions about the influence of the box size, $L_{box}$, on the gyration radius, $R_{\rm g}$, and about the coarse-grained models. We are especially interested at the predictions of the different coarse-grained models for the structure of the chain as a function of the number of monomers.
\section{System and methods}
\label{method}
\subsection{Atomistic molecular dynamics}

\begin{figure}[ht]
\centering
  \includegraphics[height=3cm]{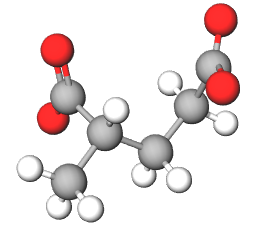}\qquad
  \includegraphics[height=2cm]{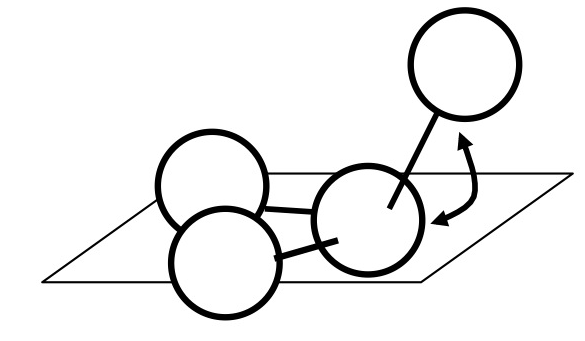}
  \caption{\emph{Left:} polyacrylate oligomer made of 2~monomers (the sodium couterions are not indicated). 
  \emph{Right:} schematic view of an improper angle between four atoms.}
  \label{fig:PAA}
\end{figure}

The  structure of an oligomer of polyacrylate made of 2~monomers is given on the left part of Figure~\ref{fig:PAA}). 
Each carbon atom bearing the carboxylate group is chiral so that a polymer of given size has several stereoisomers. Commercial NaPA chains are generally atactic, i.e. the absolute configurations of the chiral centers is randomly distributed along the chain.  

We used the same force field as described in Reith {\em et al.},\cite{Reith_JCP_2002, Biermann_these} based on the GROMOS force field\cite{Oostenbrink_WP_2004, Schuler_JCC_2001} for atomistic molecular dynamics simulations of aqueous solutions of sodium polyacrylate, and the SPC/E model\cite{Berendsen_1981} for water molecules. In GROMOS, that is a united atom force field, the hydrogen atoms of the PA chain are included in CH, CH$_2$ and CH$_3$ super atoms. With these force fields for water molecules and the NaPA, Reith {\em et al.} obtained results in agreement with experiments. Very recently, several atomistic force fields for NaPA aqueous solutions were compared, and it was shown that the GROMOS force field with SPC/E water yield results close to those obtained with an all atom force field with partial charges deduced from ab initio calculations\cite{MintisJPCB2019}. In what follows, we refer to our numerical simulations based on the GROMOS force field as atomistic simulations, even if hydrogen atoms of the PA chain are not explicit. 
In addition to usual binding, bending, and dihedral angle interactions, this force field includes an harmonic potential on the improper dihedral angles of the polyelectrolyte, $V_{\rm improper}(\delta) = K_\delta(\delta-\delta_0)^2$ (this contribution apparently was not used in Reith {\em et al.} \cite{Reith_JCP_2002}). The geometric definition of the improper angle is the same as that of the dihedral angles, but the four atoms are not consecutive. In our case, this potential applies to the elements C and O of the groups -CH--(COO$^-$) which form a "fork" on the aliphatic chain. The interaction potential forces the four atoms \textbf{C}H--(\textbf{O}O)--(\textbf{O}O)--(\textbf{C}OO) to stay in the same plane ($\delta_0 = 0$).
This potential also applies to angles \textbf{C}H$_{2/3}$--(\textbf{C}COO)--(\textbf{C}H$_{2/3}$)--\textbf{C}$^{\star}$H,
that ensures the conservation of the absolute configurations of C$^{\star}$. The parameters of the binding potentials of NaPA are collected in table~\ref{tab:paramFF_TA}. The parameters of the 12-6 Lennard-Jones potentials are gathered in table \ref{tab:paramLJ_TA}. The long range electrostatic interactions are calculated using an Ewald summation \cite{ewald}
with a cutoff radius of $12$~\AA. The carboxylate group  carries an overall charge of $-1~e$ with a charge $+0.27~e$ on the carbon atom and $-0.635~e$ on both oxygen atoms.

\begin{table}[h]
\small
\caption{Parameters of the intramolecular interaction potentials \cite{Biermann_these,Oostenbrink_WP_2004, Schuler_JCC_2001} for the polyacrylate chain in atomistic simulations.}\label{tab:paramFF_TA}
\begin{tabular}{ccccc}
\hline
\textbf{Binding}  & & & \\
Type & $K_r$ (kJ mol$^{-1}$~\AA$^{-2}$) & $r_0$ (\AA) & \\
\hline
\textbf{C}H-(\textbf{C}OO)$^-$ & 1 673.6 & 1.530 &\\  
\textbf{C}H$_n$-\textbf{C}H$_{1/2}$ &1 673.6 & 1.530 &\\  
\textbf{C}-(\textbf{O}O) & & 1.250 &  \\ 
\hline
\textbf{Angles} &  & &  \\
Type & $K_\theta$ (kJ mol$^{-1}$) & $\theta_0$ (deg) & \\
\hline
CH$_2$-CH-CH$_2$ & 260.0 & 109.5 & \\
CH-CH$_2$-CH & 265.0 & 111.0 & \\ 
\textbf{C}H$_2$-\textbf{C}H-(\textbf{C}OO) & 260.0 & 109.5 & \\
\textbf{C}H-(\textbf{C}OO)-(\textbf{O}O) & 317.5 & 117.0 & \\ 
(\textbf{O}O)-(\textbf{C}OO)-(\textbf{O}O) & 385.0 & 126.0 & \\
\hline
\textbf{Dihedral} &  & &  \\
Type & $K_\tau$ (kJ mol$^{-1}$) & $n$ & $\tau_0$ (deg) \\
\hline
CH$_3$-CH-CH$_2$-CH & 5.86 & 3 & 0  \\ 
CH-CH$_2$-CH-CH$_2$ & 5.86 & 3 & 0 \\
CH$_2$-CH-CH$_2$-CH & 5.86 & 3 & 0 \\
\textbf{C}H$_{2/3}$-\textbf{C}H-(\textbf{C}OO)-(\textbf{O}O) & 1.000 & 6 & 30.0 \\
\hline
\textbf{Impropers} &  & & \\
Type & $K_\delta$ (kJ mol$^{-1}$ deg$^{-2}$) & $\delta_0$ (deg) &  \\
\hline
\textbf{C}H-(\textbf{O}O)-(\textbf{O}O)-(\textbf{C}OO) & 0.0255 & 0 &  \\ 
\textbf{C}H$_{2/3}$-(\textbf{C}OO)-\textbf{C}H$_{2/3}$-\textbf{C}H & 0.0510 & 35.26 & \\ 
\hline
\end{tabular}

\end{table}

\begin{table}[h]
\small
\caption{Parameters of non-binding interaction potentials of the atomistic force field \cite{Biermann_these,Oostenbrink_WP_2004, Schuler_JCC_2001}}\label{tab:paramLJ_TA}
\begin{tabular}{ccccc}
\hline
Type  & $\varepsilon$ (kJ mol$^{-1}$) & $\sigma$ (\AA) & $q$ ($e$) & $m$ (g mol$^{-1}$)\\
\hline
\textbf{PA} & & & & \\
CH & 0.3139 & 3.8004 & 0.0000 & 13.0190 \\
CH$_2$ & 0.4896 & 3.9199 & 0.000 & 14.0270 \\
CH$_3$ & 0.3875 & 7.3227 & 0.000 & 15.0350 \\
\textbf{C}OO$^-$ & 0.4059 & 3.3611& +0.2700 & 12.0100 \\
C\textbf{OO}$^-$ & 1.7254 & 2.6259 & -0.6350 & 15.9994 \\
Na$^+$ & 0.3580 & 2.7300 & +1.0000 & 22.9898 \\
\hline
\textbf{water (SPC/E)} & & \\
HW & 0.0000 & 0.0000 & +0.424 & 1.0000 \\
OW & 0.6503 & 3.1650  & -0.848 & 16.0000 \\ 
\hline
\end{tabular}
\end{table}

For each atomistic system, the simulation procedure was  the following. First, an initial configuration of one PA chain and its counterions  in a box of water molecules was generated. Our reference atomistic simulations deal with a chain of $23$ monomers as in Reith {\em et al} \cite{Reith_JCP_2002} and in Mintis {\em et al}\cite{MintisJPCB2019}.
 A first equilibration in the $NVT$ ensemble allowed us to reach a state corresponding to the desired temperature (here $300$~K) from the initial configuration which was usually far from equilibrium. Then, a second series of equilibration in the $NPT$ ensemble was performed under  an ambient  pressure of $1.0$~bar and a temperature of $300$~K for $0.5$~ns. The time step was $0.5$~fs and Nos\'e-Hoover algorithms were used to control the temperature and the pressure with relaxation times  of $1$~ps each. After these two first steps, six systems of different atactic configurations of one PA chain of $23$ monomers have been obtained, containing between $3161$ and $3193$ water molecules for a box size between $45.84$ and $45.52$~\AA.  Then, trajectories in the $NVT$ ensemble with a time step of $0.5$~fs were produced for about $200$~ns in each case. Morever, to ensure that equilibrium states were reached, we have performed $3$ to $5$ independent trajectories for each atactic configuration. Indeed, it was shown previously for atomistic simulations of another polyelectrolyte in water that the conformational properties of such chains might not be sampled in a single trajectory \cite{ParkJPCB2012}. 
 
\subsection{Methodology to infer the parameters of the coarse-grained models}

From a theoretical point of view, the knowledge of all the pair correlation functions allows one to define the effective pair potential between sites of a given model\cite{Henderson74,Chayes84}. In isotropic systems, radial distribution functions are often used as input data to infer model parameters. In this study, simulations of PA are used to generate these data. More precisely, radial and angular distribution functions are obtained from atomistic simulations of one PA chain of $23$ monomers in water, in a simulation box of about $46$~\AA~(the precise value depends on the stereoisomer studied). These distribution functions are averaged over several atactic configurations, so that the resulting model is suitable to model a mixture of stereoisomers. The distributions are then used as targets for the optimization of the CG force field. The principle of this optimization is to vary systematically the parameters of a given CG model until the chosen target is correctly reproduced. 

To compute the properties of the CG models, we performed Langevin dynamics simulations \cite{padro} of a CG chain of $23$ beads in a simulation box of $46$~\AA, with the LAMMPS package\cite{lammps}. We used a time step of $10$~fs and friction coefficients equal to $3.7\times 10^{-2}$~fs$^{-1}$ for the monomers, $8.2\times 10^{-2}$~fs$^{-1}$ for the sodium ions when they are accounted for.

\subsection{Coarse-grained models}

The simplest coarse-grained (CG) model we have studied is made of neutral superatoms, as in Ref.\cite{Reith_JCP_2002}. Each site represents a chemical monomer and its counterion, and is centered on the center of mass of the chemical monomer. This model hence describes both water and counterions as implicit particles and is denoted by ImpCG model in what follows. Interactions between superatoms contain in this case a bonding contribution (spring) between beads, an angular spring between three consecutive beads, and a 12-6 Lennard-Jones interaction between beads separated by at least two other beads (we refer to this contribution as 1-4 LJ in what follows). The six parameters of the ImpCG model have been inferred by fitting the full radial distribution function between beads, using an automatic optimization algorithm (simplex \cite{Meyer_JCP_2000,Reith_M_2001}). The resulting parameters of the ImpCG model are collected in Table~\ref{tab:paramGG}. It should be noted that coarse-grained models with a Lennard-Jones attraction between bonded beads, or between beads separated by two bonds, did not lead to satisfactory sets of parameters (either the bond parameters or the angular ones obtained after the simplex optimization were aberrant in these cases), so that we have restricted the LJ attraction to beads separated by at least three bonds (1-4 LJ).
The radial distribution function between monomers obtained from the atomistic simulations is displayed in Fig. \ref{fgr:gr-full} in black, and the one obtained for the ImpCG model is displayed in blue on the same figure. Fig. \ref{fgr:gr-angular} displays the angular distribution functions between monomers with the same color code. In Sec.~\ref{rdf-ta-GG}, we provide a detailed discussion of these distribution functions.

We have also studied another CG model that explicitely accounts for the charge of the carboxylate group and for sodium counterions. This model is called ExpCG in what follows. Water is described implicitely by a continuous medium characterized by its dielectric constant.   In this case, the superatoms are still centered on the center of mass of the monomer, but they bear a central charge equal to $-1$~$e$. Sodium ions have a central positive charge $+1$~$e$. The electrostatic interactions between bonded beads, and between beads separated by two bonds were not taken into account. The dielectric constant of the SPC/E model of water does not correspond to that of real water. Nevertheless, we have chosen to fix the dielectric constant of water of our CG model at its experimental value in ambient conditions ($\varepsilon_r = 78.5$), so that the parameters we obtain after the fitting procedure are consistent with this value. 

From a numerical point of view, for a given CG model, we have noticed that several sets of parameters lead to similarly good agreement with the atomistic radial distribution functions. Our strategy was then to keep all coarse-grained models sufficiently close to each other to unravel the role of their different contributions on the average properties. 
The ExpCG model derives thus from the implicit counterion model ImpCG. It contains the same bonding, angular bonding and 1-4 LJ contribution as the ImpCG model, in addition to electrostatic interactions. The non electrostatic interactions between the monomer and its counterions are accounted for by a 12-6 Lennard-Jones interaction potential. As the beads are charged, we expect an electrostatic repulsion between distant beads, that should be partly compensated by the screening due to sodium counterions. It should be noted that electrostatic repulsions between two successive beads or two beads separated by two bonds are not accounted for. For the Na$^+$/Na$^+$ interaction, we included  the electrostatic repulsion only. 
The inferred parameters of the ExpCG charged model thus concern the monomer/Na$^+$ interaction (12-6 Lennard-Jones). The two LJ parameters were fitted so that the distribution function between beads, first nearest neighbors excluded, was correctly described. In summary, the ExpCG potential contains the influence of electrostatic interactions both in the LJ monomer/monomer contribution, and in the explicit electrostatic interactions between monomers and between monomers and counterions. The added Lennard-Jones contribution to the monomer/counterions interaction allows us to recover the correct distribution function between monomers.

\begin{figure}[ht]
\centering
  \includegraphics[height=7cm]{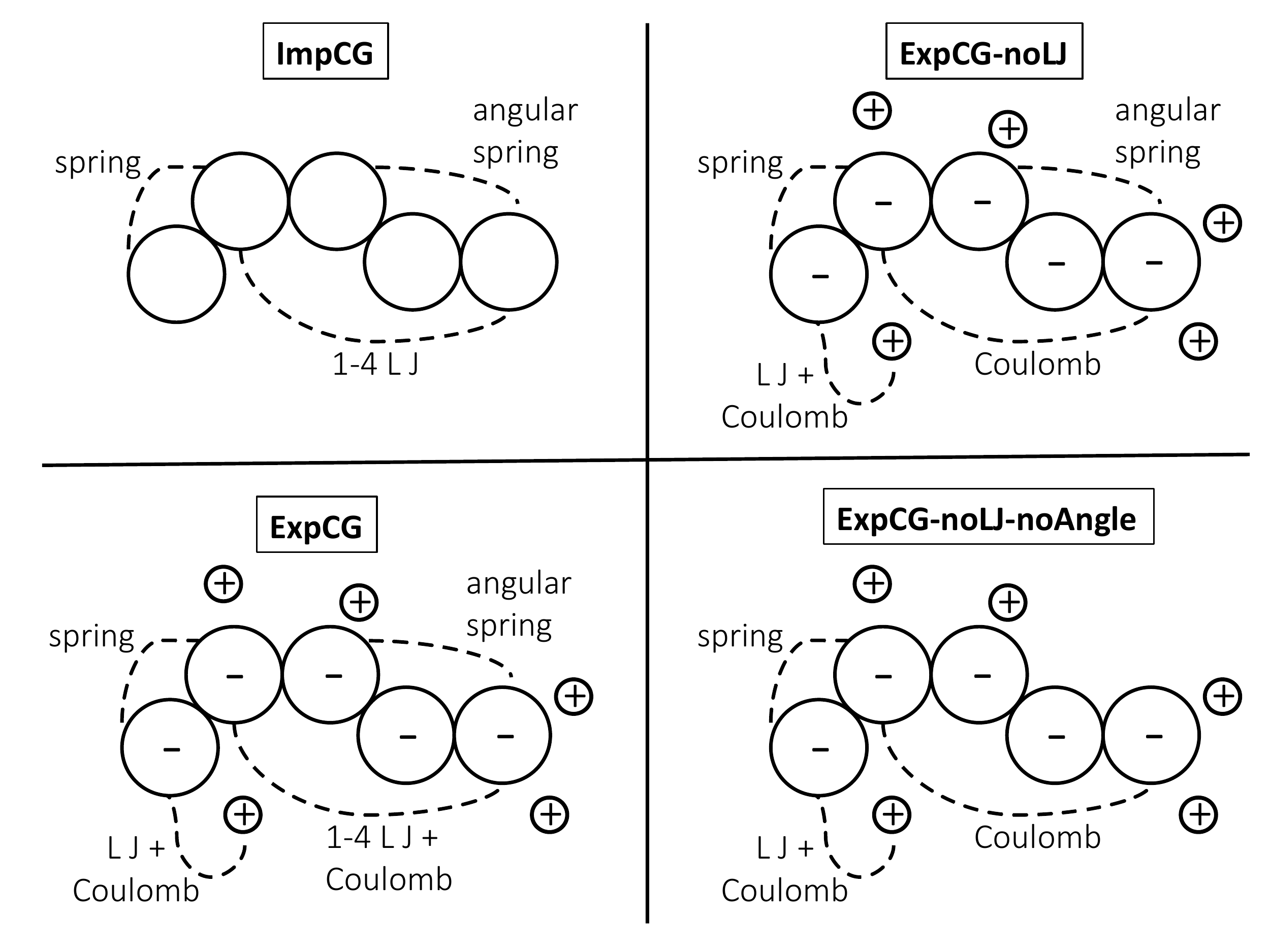}
  \caption{Summary of the different CG models emphasizing the different contributions to the interactions between superatoms.}
  \label{CGmodel}
\end{figure}

 As usual simple models of polyelectrolytes used in the literature 
 do not include attractive Lennard-Jones interactions between distant monomers, but only short-range repulsions\cite{stevens95,WinklerPRL98,Chang2002,cabaneSoftMatt2018}, we have also studied another variant of the explicitely charged coarse-grained model, called ExpCG-noLJ model. In this case, we kept the bond,  bend and electrostatic interactions of the ExpCG model, but we replaced the 1-4 LJ potential by a short ranged purely repulsive  Weeks-Chandler-Andersen (WCA) interaction potential between distant beads, with a size parameter equal to the bond length. This will allow us to study the impact of the 1-4 LJ contribution on the average properties of the PA chain. Finally, to investigate the influence of the angular contribution to the CG interaction potential, we have also studied another variant of the ExpCG model, called ExpCG-noLJ-noAngle: It contains the same interaction potentials as ExpCG-noLJ except the angular potential that is completely suppressed. ExpCG-noLJ-noAngle thus models a kind of freely jointed chain of charged beads with excluded volumes, that is very close to the generic models of polyelectrolytes used in the literature\cite{stevens95,WinklerPRL98,Chang2002,cabaneSoftMatt2018}.  ExpCG-noLJ and ExpCG-noLJ-noAngle models allow us to quantify the influence of the different types of interactions present in the ExpCG model. The parameters of explicit charged CG models are also collected in Table \ref{tab:paramGG}.
A schematic summary of the different CG models is given on Fig. \ref{CGmodel}.

 
 Finally, we have also studied the influence of an added salt, NaCl, in coarse-grained simulations. When chloride ions are present, they interact with charged beads through the electrostatic repulsion only. The Na$^+$/Cl$^-$ interaction contains the electrostatic attraction plus a short ranged repulsion (WCA interaction potential), that allows us to fix the minimal distance of approach between sodium and chloride at $3.67$~\AA. This distance is usually used to account for the structural and dynamical properties of aqueous solutions of sodium chloride described by the primitive model of electrolytes\cite{Dufreche05}.

\begin{table}[h]
\centering
\caption{Interaction parameters of the coarse-grained models. The binding contribution is $V_{\rm bond}(r) = K_r(r-r_0)^2$, the angular one is $V_{\rm angle}(\theta) = K_\theta\left[\cos(\theta)-\cos(\theta_0)\right]^2$. All models contain the same binding contribution. All charged models contain the same Bead/Na$^+$ and Na$^+$/Na$^+$ contributions. ExpCG-noLJ-noAngle has no angular potential. 1-4 LJ denotes a 12-6 interaction potential between beads separated by at least two other beads. ExpCG* stands for "ExpCG, ExpCG-noLJ, ExpCG-noLJ-noAngle", ExpCG-noLJ* stands for "ExpCG-noLJ, ExpCG-noLJ-noAngle".}\label{tab:paramGG}
\begin{tabular}{cccc}
\hline
\textbf{Binding}  
 & $K_r$ (kcal mol$^{-1}$~\AA$^{-2}$) & $r_0$ (\AA) & \\

& 1.6872 & 3.5288 \\
\hline
\textbf{Angles} & $K_\theta$ (kcal mol$^{-1}$) & $\theta_0$ (deg) & \\

 & 2.5268 & 86.1042 & \\

\hline
\textbf{1-4 LJ} &  $\varepsilon$ (kcal mol$^{-1}$) & $\sigma$ (\AA)\\
ImpCG,ExpCG & 0.2249 & 4.5160 \\
ExpCG-noLJ* & 1.42 &  3.53 \\
\hline
\textbf{Bead/Na$^+$} &  $\varepsilon$ (kcal mol$^{-1}$) & $\sigma$ (\AA)\\
ExpCG* & 0.50 & 2.75 \\
\hline
\textbf{Na$^+$/Cl$^-$} &  $\varepsilon$ (kcal mol$^{-1}$) & $\sigma$ (\AA)\\
ExpCG* & 0.026 &  3.67\\
\hline
\end{tabular}
\end{table}

\subsection{Structural properties of the PA chain}

Once the CG models are determined, their ability to describe the structure of the system and its evolution are investigated. The structure of the chain  is mainly characterized through the distribution over time and the average value of the gyration radius.
Indeed, a standard way to quantify the mean size of a single polymer chain in a given configuration is the standard deviation of its position distribution, or the radius of gyration $R_{\rm g}$:
\begin{eqnarray} \label{eq:RG1}
	R_{\rm g}^2
	    &= \frac{1}{N}\sum_{i=1}^{N} \left( {\bf r}_{i} - {\bf  r}_G \right)^2 \nonumber \\
	   &  = \frac{1}{N}\sum_{i=1}^{N} ({\bf r}_i^2 - {\bf  r}_G^2), \nonumber \\
	&\quad \mathrm{with} \quad {\bf  r}_G = \frac{1}{N}\sum_{i=1}^{N} {\bf  r}_i \,
\end{eqnarray}
$N$ being the number of monomers of the chain, and ${\bf r}_i$ the position of the $i$-th monomer.

It is common to statistically characterize the average behavior of a polymer of $N$ monomers by means of the mean radius of gyration,
\begin{equation} \label{eq:meanRG}
	\overline{R}_{\rm g} = \sqrt{\left\langle {R}_{\rm g}^{2}\right\rangle} \,,
\end{equation}
where the average $\langle \cdot \rangle$ is performed over the ensemble of conformations for a given polymer. The \emph{scaling behavior} of $\overline{R}_{\rm g}$ with the polymer length $N$ 

reads
\begin{equation}
\overline{R}_{\rm g} \propto (N-1)^\nu
\end{equation}
where the scaling exponent $\nu$ is the so-called  \emph{Flory exponent}\cite{Flory_1953,DeGennes1979} (see section \ref{scaling}).

The hydrodynamic radius of a polymer is the inverse of the average distance between any monomers $i$ and $j$:
\begin{equation}
    	\frac{1}{\overline{R}_{\rm h}} = \frac{1}{N^2}\sum_{i,j=1,i\ne j}^{N}\left\langle\frac{1}{\sqrt{ \left( {\bf r}_{i} - {\bf  r}_{j}\right)^2}}\right\rangle.
\end{equation}

\section{Results and discussion}
\label{results}
\subsection{Radial distribution functions from atomistic simulations and coarse-grained models for the chain of 23~monomers}
\label{rdf-ta-GG}

\begin{figure}[ht]
\centering
  \includegraphics[height=7.5cm]{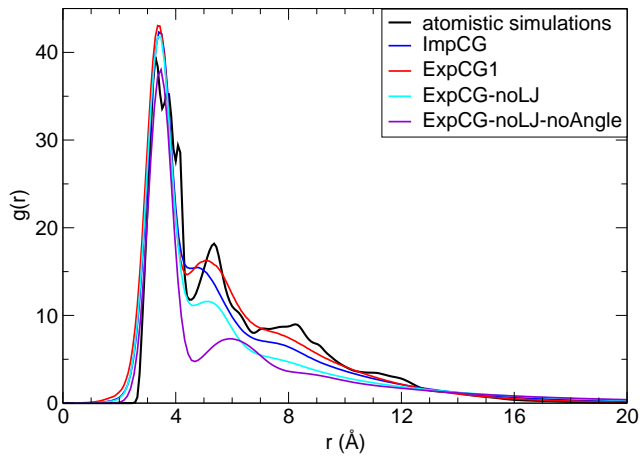}
  \caption{Radial distribution functions between the centers of mass of monomers of the PA chain for a chain of $23$ monomers in a simulation box of $46$~\AA.}
  \label{fgr:gr-full}
\end{figure}

\begin{figure}[ht]
\centering
  \includegraphics[height=7.5cm]{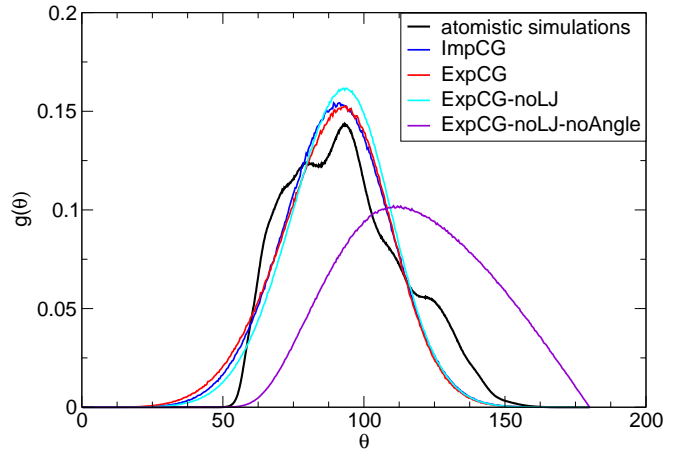}
  \caption{Angular distribution functions between three successive the centers of mass of monomers of the PA chain, first neighbors excluded,  for a chain of $23$ monomers in a simulation box of $46$~\AA. }
  \label{fgr:gr-angular}
\end{figure}

\begin{figure}[ht]
\centering
  \includegraphics[height=7.5cm]{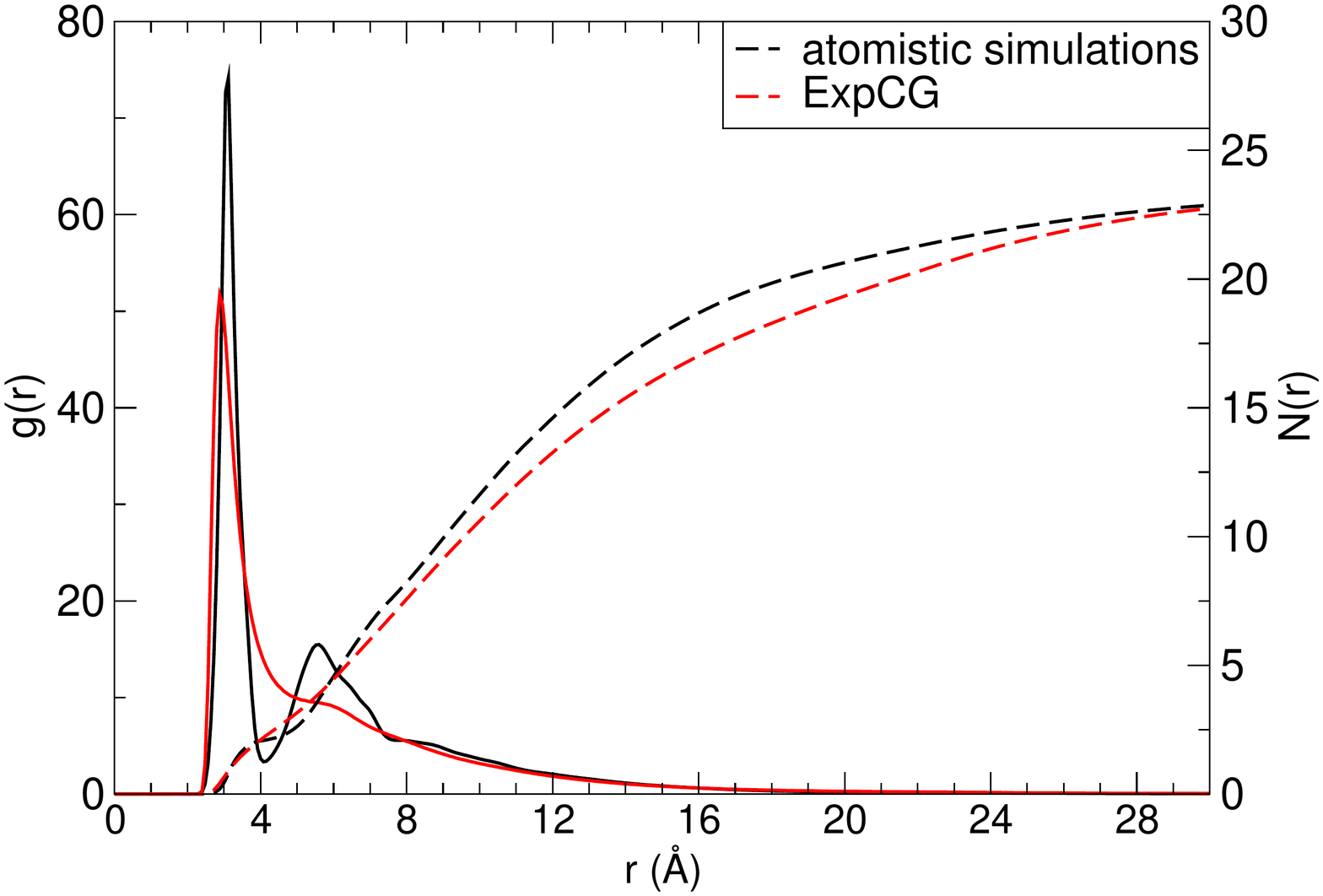}
  \caption{ Radial distribution functions between the centers of mass of monomers of the PA chain and the counterions Na$^+$ for a chain of $23$ monomers in a simulation box of $46$~\AA. In dashed lines the corresponding coordination numbers $N(r)$, obtained from an integration of the radial distribution functions are given. As expected, $N(r)$ tends to the total number of counterions in the simulation box, $23$, at long distance.}
  \label{fgr:gr-mon-cti}
\end{figure}

The radial and angular distribution functions between monomers computed from atomistic simulations and for the coarse-grained models are shown in Figs. \ref{fgr:gr-full} and \ref{fgr:gr-angular}. The results of atomistic simulations plotted here are averages over the six different atactic configurations. We recall that the parameters of ImpCG and ExpCG models were fitted to account for these distribution functions. As it can be seen on these figures:
\begin{itemize}
     \item All CG models correctly reproduce the first peak of the radial distribution function (rdf) that corresponds to the monomer-monomer bond. This was expected as the parameters of the ImpCG model were optimized to reproduce this radial distribution function, and as the bonding parameters are those of ImpCG for all the CG models.
     \item The atomistic simulations predict a clear second peak of the rdf at about $5$~\AA. All rdfs obtained using CG models present a second peak. ImpCG and ExpCG correctly reproduce the average intensity of this second peak of the rdf between monomers, although the shape of the peak is less sharp. The intensity of the peak is closer to the atomistic result when explicit charges are described (ExpCG versus ImpCG). On the other hand, ExpCG-noLJ and ExpCG-noLJ-noAngle, whose parameters were not fitted to recover the atomistic results, except the bonding and angular potentials, fail to describe correctly the second peak of the radial distribution function. When Lennard-Jones attractions between distant monomers are switched off, the CG models cannot predict correctly the intensity of the second peak. If angular constraints are added (see ExpCG-noLJ), the second peak is closer to the target, but these constraints are not sufficient to mimic the influence of the LJ attraction.
     \item The third peak of the atomistic distribution function is not correctly reproduced by CG models. ExpCG-noLJ and ExpCG-noLJ-noAngle yield similar distribution functions for distances larger than $8$~\AA, i.e. the influence of the angular constraint is no longer visible for large distances. This makes sense since this peak does not correspond to second neighbors and is not directly affected by the angular potential. 
     \item The shape of the angular distribution function obtained from atomistic simulations is not perfectly Gaussian and symmetric. This feature is absent in all coarse-grained models, where the angular constraint is harmonic and yield Gaussian angular distributions. All CG models correctly reproduce the mean position of the peak of the angular distribution function, except the ExpCG-noLJ-noAngle as it does not include an angular interaction potential. The angular distribution in this case is only influenced by long-ranged electrostatic interactions between monomers. This distribution is also broader, which is consistent with the greater extension of chains obtained by this model (see hereafter). 
\end{itemize}
We display in Fig.  \ref{fgr:gr-mon-cti} the radial distribution functions between monomers and counterions, as well as the coordination number $N(r)$, that is the total number of counterions at a distance $r$ of a given monomer. As it can be seen on this figure, the position of the first peak of the $g(r)$ is correctly described by the CG model, but the peak is larger. The $g(r)$ obtained from atomistic simulations presents a well defined second peak, that is only rendered by a shoulder at the coarse-grained level. Nevertheless, the coordination numbers coincide for a distance equal to $8$~\AA, that corresponds to the end of the second peak of the atomistic $g(r)$. This shows that our fitting procedure, that aimed at reproducing the monomer-monomer distribution functions, allows us to describe adequately the distribution function between monomers and counterions.
  
\subsection{Evolution of the polymer size in atomistic simulations of different atactic configurations}

\begin{figure}[ht]
\centering
  \includegraphics[height=7.5cm]{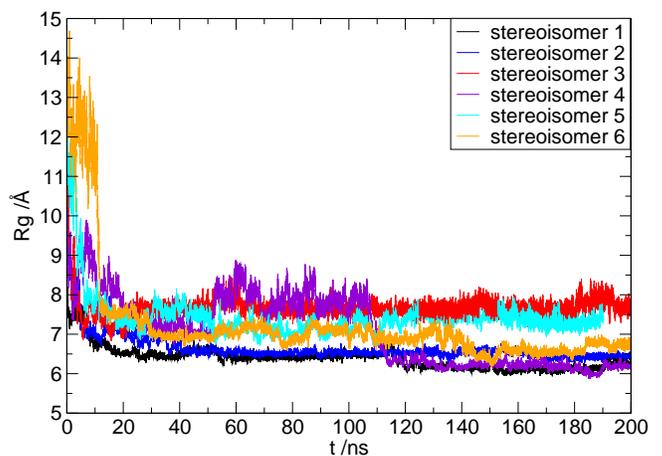}
  \caption{Gyration radius of a chain of $23$ monomers in a simulation box of $46$~\AA~ as a function of time obtained from atomistic simulations for six different atactic configurations, starting from extended configurations of the chain.}
  \label{fgr:Rg-AA}
\end{figure}

\begin{figure}[ht]
\centering
  \includegraphics[height=7.5cm]{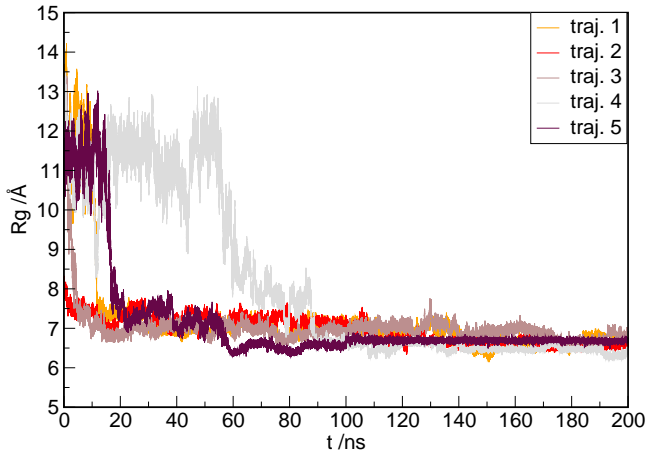}
  \caption{Gyration radius of one atactic configuration (stereoisomer $6$) of  a chain of $23$ monomers in a simulation box of $46$~\AA~ as a function of time obtained from atomistic simulations for five independent trajectories. Note that the trajectory number $1$ (orange line) is the same as that shown on Fig. \ref{fgr:Rg-AA}, also as an orange line.}
  \label{fgr:Rg-atac6}
\end{figure}

\begin{figure}[ht]
\centering
  \includegraphics[height=7.5cm]{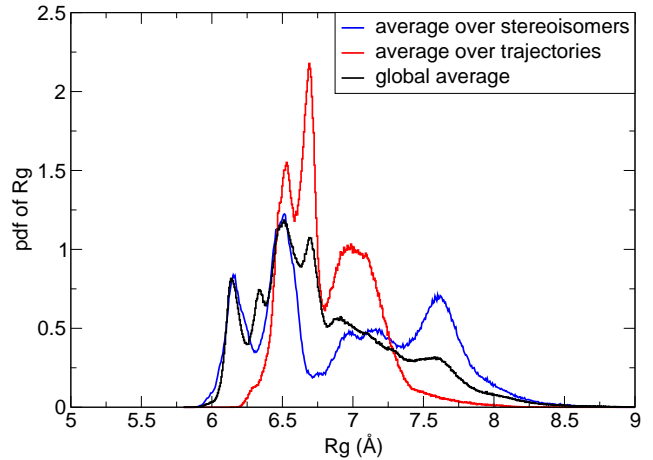}
  \caption{Probability density function of the gyration radius obtained from trajectories of one chain of $23$ monomers in a simulation box of $46$~\AA. Blue: results obtained from  atomistic trajectories for six different atactic configurations (one trajectory each); Red: results obtained from  atomistic trajectories  for five independent trajectories for one atactic configuration (stereoisomer $6$); Black: results obtained from  $18$ independent atomistic trajectories  (averaged over several atactic configurations, and several trajectories for each atactic configuration).  }
  \label{fgr:histo-Rg-AA-traj}
\end{figure}

\begin{figure}[ht]
\centering
  \includegraphics[height=7.5cm]{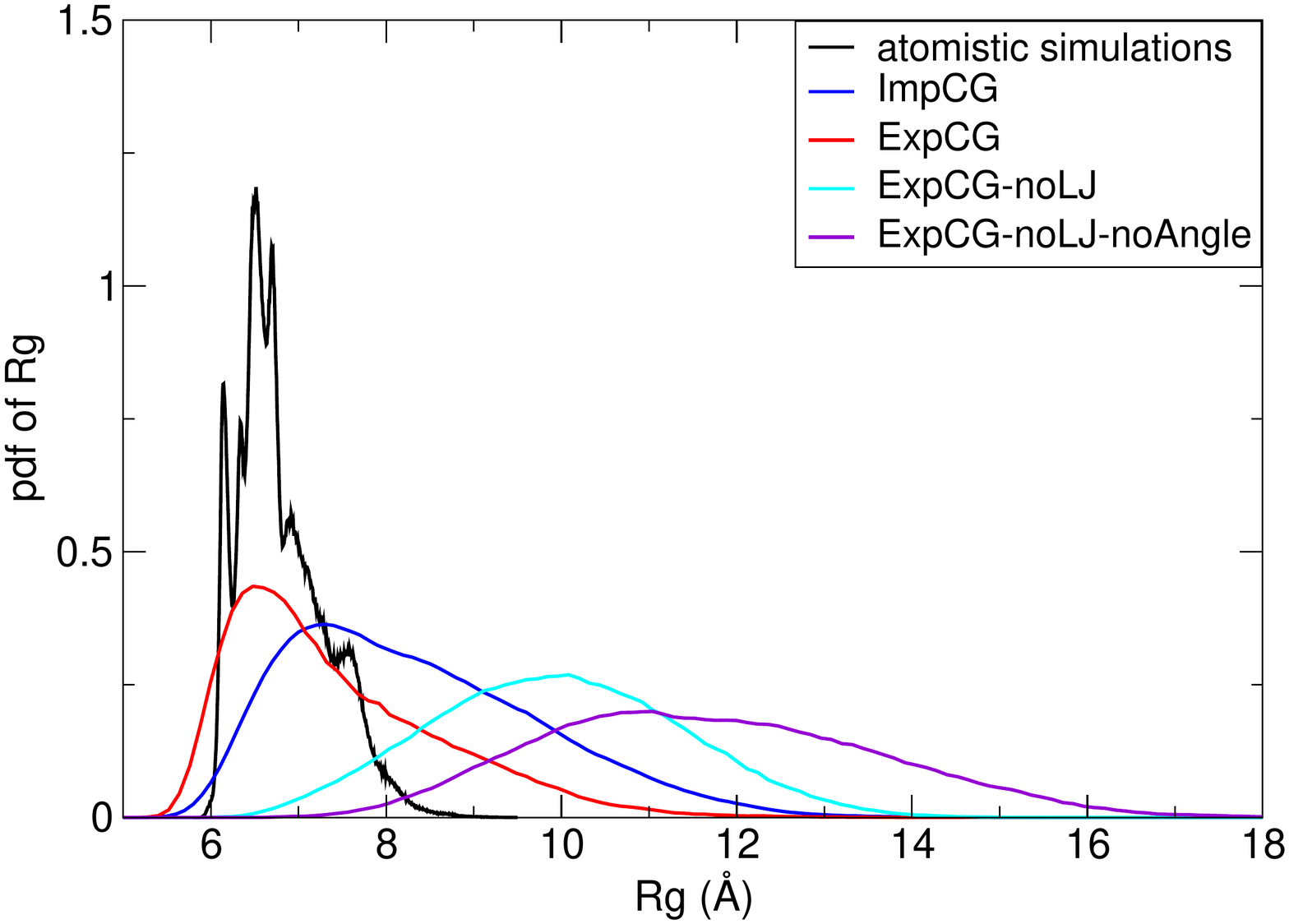}
  \caption{Probability density function of the gyration radius obtained from trajectories of one chain of $23$ monomers in a simulation box of $46$~\AA. Black: results obtained from the atomistic trajectories  between $25$ and $200$~ns for six different atactic configurations and several independent trajectories for each atactic configuration (this plot is the same as the black on in Fig. \ref{fgr:histo-Rg-AA-traj}); Blue: fitted CG model with implicit counterions (ImpCG); Red: fitted model, with explicit ions (ExpCG); Cyan: ExpCG model without attractions between distant monomers and with an angular interaction potential (ExpCG-noLJ); Violet: ExpCG model without LJ attraction, without angular interaction (ExpCG-noLJ-noAngle).   }
  \label{fgr:histo-Rg-AA}
\end{figure}

Previous atomistic simulations of the same system with the same force field have been performed\cite{Reith_JCP_2002}. The improving computational capabilities enable to challenge these previous data on two aspects: the convergence over time of the internal structure of the polymer, and the relevant average over different isomeric configurations. The later point allows us to discuss the impact of the coarse-graining process, as we move from asymmetric monomers to centrosymmetric ones.

 The gyration radius of the PA chain was computed as a function of time from each atomistic molecular dynamics trajectory. The results are given in Fig.~\ref{fgr:Rg-AA} and Fig.~\ref{fgr:Rg-atac6}. In Fig.~\ref{fgr:Rg-AA}, we present the results obtained for $6$ different atactic configurations; One trajectory of $200$~ns was computed for each stereoisomer. In Fig.~\ref{fgr:Rg-atac6}, we show the results obtained for one given atactic configuration for $5$ independent trajectories. Other trajectories for other stereoisomers were also computed and are not shown here. In total, we have obtained $18$ independent trajectories of duration equal to $200$~ns each.
As shown on these figures, for a given trajectory, the time needed to reach a stationary state with a gyration radius fluctuating around an almost constant value is of the order of at least $10-20$~ns. In one case (trajectory number $4$ in Fig.~\ref{fgr:Rg-atac6}) $80$~ns were needed to reach a stationary state. Also, the stereoisomer 4 (violet curve on Fig. ~\ref{fgr:Rg-AA}) has a gyration radius that varies with time with a longer time scale.
 These results raise the question of the quality of the convergence of atomistic trajectories data. Even for a relatively long simulation corresponding to $200$~ns, slow dynamics, maybe related to the existence of metastable states, may impact the structural data collected during the simulation.     This issue was already discussed in Park {\em et al.}\cite{ParkJPCB2012} in the case of polystyrene sulfonate chains: These authors concluded that atomistic simulations of such systems  should be performed either from Hamiltonian replica exchange molecular dynamics or by averaging over many independent trajectories. In the present study, we observe that 
 different atactic configurations (stereoisomers) converge to different mean gyration radii.  As it appears clearly on Fig.~\ref{fgr:Rg-AA}, the amplitude of the fluctuations of the gyration radius varies from one stereoisomer to another at a given temperature.  Also, for a given stereoisomer, several independent trajectories may lead to different values of the gyration radius, and to stationary states with fluctuations of different amplitudes, as illustrated in Fig. ~\ref{fgr:Rg-atac6}.
 
 We display in Fig.~\ref{fgr:histo-Rg-AA-traj}  the distributions of the gyration radius computed from atomistic trajectories. Note that to compute these distributions, we have used the part of the trajectories that converged to stationary states. For example, data from trajectory $5$ of stereoisomer $6$ were used between $10$ and $200$~ns whereas data from trajectory $4$ were used between $85$ and $200$~ns (see Fig. \ref{fgr:Rg-AA}). We show in this figure the distribution obtained by averaging over several stereoisomers, with one trajectory for each (blue curve), as well as the distribution obtained by averaging over several independent trajectories for a single stereoisomer (red curve). Also, the distribution obtained by averaging over several stereoisomers and over several trajectories for each stereoisomers is plotted in black. As it appears on Fig.~\ref{fgr:histo-Rg-AA-traj}, the distribution of gyration radius contains several peaks in every case, due to both thermal fluctuations and stereoisomery. It appears that the distribution obtained for several stereoisomers (blue curve) has almost the same width as the one obtained from several trajectories of the same stereoisomer (red curve). Note that another recent simulation study\cite{Katiyar2017} has emphasized the influence of tacticity on the properties of PA chains: for three different tacticities (syndiotactic, isotactic and one atactic) of a chain of $20$~monomers, the  gyration radius varies of about $20$\%. It should also be noted that, to ensure that asymetric carbons of the chain keep their configuration during the simulation, the atomistic force field must include an harmonic potential on the improper dihedral angles (see Section~\ref{method}). 


 The average value of the gyration radius computed from the distribution obtained from all the $18$ atomistic trajectories, that is thus averaged over several atactic configurations and over several trajectories for each stereoisomer is equal to $6.9$~\AA, with a distribution of radii between $6$ and $9$~\AA. If we perform the average over the $6$ stereoisomers, keeping only one trajectory for each one, the mean gyration radius is  equal to $7.0$~\AA. For stereoisomer number $6$, averaging over five independent trajectories yields a mean gyration radius of $6.8$~\AA. 
 Finally, it appears that averaging over several stereoisomers, with only one trajectory per stereoisomer leads to almost the same mean value as averaging over several trajectories per stereoisomer. This also shows that this set of $18$ trajectories represent a good sampling of the equilibrium configuration of the PA. 
 
  The average value of the gyration radius obtained here from molecular dynamics simulations is smaller than that obtained by Reith {\em et al.}\cite{Reith_JCP_2002} at 333 K from atomistic trajectories ($R_{\rm g} = 12.8$~\AA) with a force field very close to ours, except for the potential on the improper dihedral angles that we have added,  and for the water model (SPC in Reith {\em et al.}, SPC/E in the present work).  More importantly, as previously mentioned, the trajectories in ref. \cite{Reith_JCP_2002} were short (2 ns), so that the equilibrium state might not have been reached. Also, only one atactic stereoisomer was simulated.  More recent atomistic simulations of a PA chain based on two different force fields were performed by Sulatha {\em et al.}~\cite{Sulatha_IECR_2011}, who computed the gyration radius from the last $10$~ns of $15$~ns trajectories and obtained $R_{\rm g}=10.6$~\AA~ for one atactic chain of 20~monomers at 300~K in a cubic box of $60$~\AA. Lastly, a very recent paper\cite{MintisJPCB2019} compared the gyration radius obtained from long atomistic simulations with ten different force fields, and obtained for one atactic chain of $23$ monomers (precise configuration not given) at 300 K in a cubic box of $48$~\AA~ a gyration radius between $8$ and $14$~\AA~ according to the force field. It appears thus from our work and from the literature, that the obtained gyration radius of a relatively short polyelectrolyte chain depends on the force field, on the absolute configuration of the asymmetric carbons and on the duration of the simulation. 
As we show in Section \ref{box}, in salt-free systems, the gyration radius also depends on the simulation box size. However, our goal in this paper is to compare the ability of several coarse-grained models of charged polymers to predict the structural properties of a given atomistic model. Our approach may be repeated to other atomistic models if needed, and our discussion on the ability of different classes of CG models to reproduce atomic properties should in principle be independent on the precise choice of the atomic model.

\subsection{Comparison of the values of gyration radius obtained from atomistic simulations and coarse-grained models}

Fig.~\ref{fgr:histo-Rg-AA} displays the distribution of the gyration radius obtained from different CG models and from the atomistic model. First, the coarse-graining of the chain leads to distributions of the gyration radius that are much broader than the one obtained from the atomistic description. This clearly indicates the limits of the CG models that loose the molecular details.
This being said, the discrepancy between the probability density functions of the gyration radius obtained for the coarse-grained models fitted on monomer-monomer radial distribution functions and the atomistic model is puzzling. In Interacting Self-Avoiding Walks (ISAW) models of polymers, only short ranged interactions are described, and the interaction parameter governs all properties, including the distribution of gyration radius\cite{Care2014}. 
Here, for the atomistic model, gyration radii above $9$~\AA~ are not observed, while for the ExpCG model, there is a significant part of the distribution above this threshold (about 13~\% of the integral of the pdf). This shows that the imperfect reproduction of rdfs can lead to important structural differences between the atomistic model and the best inferred CG models. Nevertheless, the folding mode of polymers is better characterized through the investigation of scaling laws relating the size of the polymer to the number of monomers, than through the shape of the size distribution\cite{DeGennes1975,Care2014}. We will investigate these scaling laws in a subsequent section. 

Important differences between gyration radii are seen when different CG models are considered. The pdf of the gyration radius obtained by the ExpCG model is significantly closer to the atomistic result than the ImpCG model, for which about 30~\% of the integral of the pdf correspond to values of gyration radius larger than $9$~\AA. Unexpectedly, explicit electrostatics favor a more folded structure. For this reason, the use of effective electrostatics should not be relevant for such system. Indeed, adding repulsive screened potentials (such as Debye-H\"uckel potentials\cite{debye,Soysa2015}) can only expand the polymer, which contradicts the influence of electrostatics observed here. We will come back to this point in the next section (\ref{box}). ExpCG-noLJ and ExpCG-noLJ-noAngle models, without attractive LJ contributions, lead to almost symmetric distributions for $R_{\rm g}$, contrarily to ImpCG and ExpCG, and are centered on larger means that ImpCG and ExpCG. It shows that the LJ attraction between distant monomers qualitatively affects the folding of the chain, even in the presence of explicit electrostatic interactions between monomers that can in principle give rise to chain collapse\cite{Brilliantov98}. 

The mean values of the radii are collected in Table \ref{tab:Rg23} and shall be compared with the mean values obtained from atomistic simulations. For the coarse-grained models, the uncertainties indicated in this Table were computed using the standard deviation of the mean $R_{\rm g}$  obtained from the different independent trajectories. Hence, they do not reflect the width of the distribution of the radii but the statistical quality of the CG simulations. To evaluate the uncertainty of the gyration radius computed by atomistic simulations, we have computed the standard deviation of $R_{\rm g}$  obtained from the $18$ independent trajectories We obtain a rather good agreement between the predictions of the ImpCG and ExpCG models and the atomistic simulations results, for these average values. The gyration radius computed from ExpCG is closer to the atomistic result than that of ImpCG. We also display in this table the mean values of the hydrodynamic radii obtained from the different models. These values are relatively close to each other, except again for the ExpCG-noLJ-noAngle model. It should be noted that this quantity is more sensible to the short distances between monomers than to the long ones, which can explain the small observed differences.

Beyond these first comparisons, corresponding to a unique system and a unique atomistic reference, the performance of the CG models will then be evaluated through their ability to predict variations of the averaged gyration radius. 

\begin{table}[h]
\centering
  \caption{Gyration radius ($R_{\rm g}$) and hydrodynamic radius ($R_{\rm h}$) computed from different models of a chain of $23$ monomers in a simulation box of $46$~\AA. These results of CG simulations are averaged over $6$ independent trajectories of $1\times 10^{-6}$~s each, obtained from Langevin simulations.   }
  \label{tab:Rg23}
  \begin{tabular}{c c  c }
    \hline
    Model & $R_{\rm g}$ (\AA) &  $R_{\rm h}$ (\AA) \\
    \hline
    atomistic & $6.9\pm0.40$  & $8.0\pm0.40$\\
    ImpCG & $8.28\pm0.08$&   $8.58\pm0.04$ \\
    ExpCG & $7.45\pm0.02$  &$8.08\pm 0.06$\\
    ExpCG-noLJ & $9.94\pm0.08$  &$9.45 \pm 0.04$ \\
    ExpCG-noLJ-noAngle & $11.68\pm0.12$ & $10.78\pm0.04$\\
    \hline
  \end{tabular}
\end{table}

\subsection{Influence of the box size on the gyration radius}
\label{box}
For salt-free systems, the effect of the box size on the polyelectrolyte structure is expected to be important\cite{stevens95,Brilliantov98}. Indeed, the electrostatic screening length is directly related to the concentration of counterions, proportional to the polymer concentration, and therefore to the volume of the simulated system. As the polymer concentration gets smaller, counterions tend to explore regions at long distances from the monomers, and the electrostatic repulsion between monomers becomes more important. This is a clear weakness of the implicit ion model ImpCG, if the strategy consists in determining the interaction potential parameters from atomistic simulations, i.e. for a given polyelectrolyte concentration. In most cases, due to computation time constraints, atomistic simulations are performed for rather concentrated systems. In our case, for the chain of $23$ monomers in a box of $46\times 46 \times 46$~\AA$^3$, we have a concentration of sodium polyacrylate equal to $c = 39 $~kg~m$^{-3}$. This value is large, although it is still much below the critical concentration $c^\star$ above which the polymer chains overlap. We can estimate $c^\star$ from the gyration radius, here $c^\star=M/[N_A (2 R_{\rm g})^3]\simeq1\times 10^3$~kg~m$^{-3}$ with $M$ the molar weight of the chain (here $M=1663$~g~mol$^{-1}$) and $N_A$ the Avogadro number.

\begin{figure}[ht]
\centering
  \includegraphics[height=7.5cm]{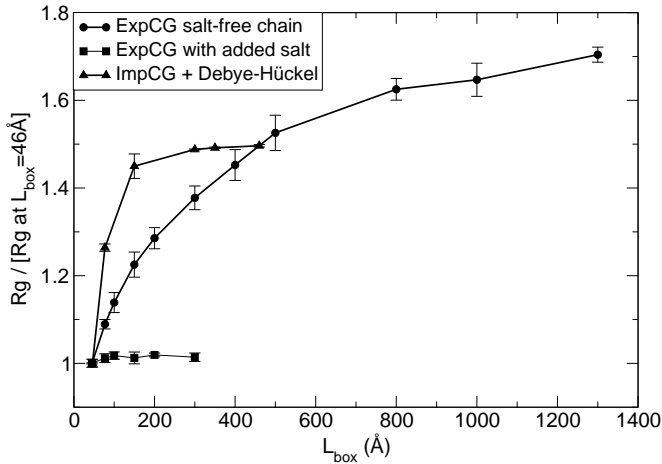}
  \caption{Average gyration radius obtained for a chain of $23$ monomers from Langevin simulation in a cubic simulation box of size $L_{\rm box}$ divided by the value obtained in the smallest simulation box, for several cases. Bullets correspond to the salt-free system modelled by the ExpCG model, i.e. with explicit electrostatic interactions. 
 Squares correspond to the ExpCG model in the case with added NaCl salt at a constant concentration equal to $0.14$~mol~dm$^{-3}$. Triangles correspond to a modified ImpCG model: screened electrostatic repulsion are added to the ImpCG model with a Debye length that depends on the box size.}
  \label{fgr:effect-box}
\end{figure}

The effect of the box size, $L_{box}$, on the average gyration radius computed for the ExpCG model is depicted in Fig.~\ref{fgr:effect-box} for a chain of $23$~monomers (black circles). The results presented here have been obtained from simulations of one single polyelectrolyte chain in the simulation box, and we checked in several cases that we obtained exactly the same results with $3$ polyelectrolyte chains in a larger box, keeping the same monomer concentration. The gyration radius has been divided by its value in the most concentrated case, here $R_{\rm g}=7.45$~\AA\ in the box of $46$~\AA. It is clear that in this regime, the influence of the monomer concentration on the gyration radius is strong. The gyration radius is found to increase with the box size: It reaches a plateau value at high dilution ($R_{\rm g}=12.7$~\AA). In this case, $c = 2\times 10^{-3}$~kg~m$^{-3}$ for $L_{\rm box}=1300$~\AA~ and $R_{\rm g}$ is about $1.7$ times larger than its smallest value, corresponding to the smallest box size ($46$~\AA) that is the same as atomistic simulations ($c = 39 $~kg~m$^{-3}$). The average gyration radius stays smaller than the value we would observe for a rod, that would be of the order of $\ell/\sqrt{12}=(22\times3.5)/\sqrt{12}\simeq 22$~\AA, where $\ell$ is the length of the rod, here $\ell\simeq (N-1)\times R_0$, with $R_0=3.5$~\AA~the mean distance between bonded monomers as deduced from the radial distribution function. It means that electrostatic repulsions between monomers are not high enough to completely extend the chain, even at the smallest concentration, for $L_{\rm box}=1300$~\AA, where the molar concentration of counterions is very small $c_{\rm Na^+}\simeq2\times10^{-5}$~mol~dm$^{-3}$).

We have also computed the average gyration radius in the presence of NaCl at a constant concentration equal to $0.14$~mol~dm$^{-3}$. In this case, the gyration radius obtained at the highest polymer concentration ($L_{\rm box}=46$~\AA) is slightly smaller than the value in the salt-free system: $R_{\rm g}=7.14$~\AA~ with salt instead of $R_{\rm g}=7.45$~\AA~in the salt-free system. Moreover, as it is shown in Fig.~\ref{fgr:effect-box} (filled squares), the gyration radius becomes independent of the polyelectrolyte concentration in the presence of added salt. Indeed, the electrostatic screening length is almost constant when the box size changes, as it is mainly governed by the added salt concentration. When considering only the added salt, the Debye screening length is equal to about $8$~\AA~in this case, i.e. almost twice the monomer-monomer bond length in the CG model. It suggests electrostatics is fully screened in this regime, so that interactions between monomers that are far away along the chain are mainly contact interactions, such as in the ImpCG model.  

We obtained the same trend with all CG models with explicit ions: In the presence of added salt no significant box size effect is observed. The gyration radius is actually similar to that of a small simulation box with $L_{\rm box}=40$~\AA, i.e. at a relatively large concentration. In their previous work, Reith et al calibrated their implicit ion CG models on a salt-free atomistic system at relatively high concentration, and compared the predictions of the CG models with experimental results at high salt concentration. They were fortunate to have chosen an atomistic situation of a salt-free polyelectrolyte with a screening length equivalent to that of a solution at high salt concentration.

The ImpCG model, implicit {\em neutral} model, is justified as long as effective electrostatic repulsions between beads are not too important. Conversely to the colloidal sciences, the use of implicit screened electrostatics for linear polyelectrolytes is very rare. Indeed, the decomposition of the effective interactions into Yukawa interactions between monomers is not accurate. Nevertheless, as a matter of completeness, we designed a second implicit counterion model, using the ImpCG model to which we added a screened electrostatic repulsion between monomers using Debye-H\"uckel interactions\cite{debye,YethirajJCP98,Soysa2015}. One may indeed wonder whether the use of effective electrostatics interactions at the Debye-H\"uckel level would allow us to account for the effect of the box size on the gyration radius of the salt-free chain of $23$ monomers. To compute the Debye length $\kappa^{-1}$, that is the screening parameter of this interaction potential for this series of simulations, we have used only the counterions of the PA chain. For $L_{\rm box}=46$~\AA~ we have $\kappa^{-1}\simeq4$~\AA, and for $L_{\rm box}=450$~\AA~we have $\kappa^{-1}\simeq150$~\AA. For the most concentrated system, we obtained $R_{\rm g}=10.8$~\AA~with screened electrostatic repulsions added to the ImpCG model. This value is of course larger than that obtained with ImpCG because we have added a repulsion. In Fig. \ref{fgr:effect-box}, we have thus plotted the evolution of the gyration radius rescaled by its value in the most concentrated case.
As with explicit electrostatics, as it appears on Fig.~\ref{fgr:effect-box}, the gyration radius significantly increases with the box size (filled diamonds). Nevertheless,  the dependence on the box size is very different with the Debye-H\"uckel model. First, the increase of $R_{\rm g}$ is stronger than with ExpCG, and second, the relative increase is smaller than with ExpCG. This result shows that the dependence of the size on the polymer concentration is not well captured by screened electrostatics. 

Taken together, these results point out the risk in using a coarse-grained model which does not explicitely include charges, or with a too crude treatment of implicit electrostatics (Debye-H\"uckel level). For protein simulations for instance, it is usual to use a quite sophisticated treatment of electrostatics at the Poisson-Boltzmann level, and in this case the relevant Debye length is used as input. However, this kind of calculation is relatively expensive, even compared with the more accurate explicit representations of small ions we have here.  

\subsection{Transferability of the coarse-grained models to other chain lengths}

 Scaling of polymer properties with the number $N$ of monomers is at the heart of polymer physics. Scaling laws emerge from the scale invariance of polymers, which are fractal objects. However, such scale invariance breaks for small chains\cite{Dolce2017}, due to finite size effects that can alter the properties of the polymer. Moreover, polymer chains often show a different scaling regime at different scales, for instance a polymer can behave as an ideal chain at small length scale and as a condensed globule at a larger scale\cite{Grassberger1995,Care2014}. In this case, the scale boundary corresponds to the definition of a blob. A key question in polymer coarse-graining is to determine whether scaling properties, deviations from scaling laws, or a change in the scaling regime can be quantified in the microscopic (atomistic) model, and reproduced in the mesoscopic (coarse-grained) model.
 
 The previous study by Reith {\em et al.} focused on the scaling properties for long chains (until $N\simeq 3000$), in the presence of added salt at a large concentration ($c_{\rm salt} > 0.14$~mol~dm$^{-3}$). In this regime, they obtained a scaling regime corresponding to a self-avoiding walk ($\nu=0.58$), in agreement with experimental data from dynamic light scattering. This means that the electrolyte nature of the chain is not visible at this scale, i.e the PA behaves as a neutral polymer. 
  Unfortunately, experimental data are not available for smaller chains due to low scattering. One can nevertheless expect that at shorter length scales, finite size effects become important and might reveal the impact of electrostatics into short range structural correlations.

We focus in what follows on two kinds of comparisons : (i) for short chains, we compare the predictions of the CG models to the results of atomistic simulations, (ii) for longer chains, we compare the predictions of the CG models with each other, in salt-free solutions and with added salt, for series of simulations at fixed total monomer concentrations.  We have kept the CG parameters fitted for the chain of $23$ monomers (see Table  \ref{tab:paramGG}) in what follows. 
 
\subsubsection{Comparison to atomistic simulations of other chain lengths}

\begin{table}[h]
\centering
  \caption{Gyration radius (\AA) computed from different models of PA oligomers.  $N$ denotes the number of monomers. The total monomer concentration varies from one value of $N$ to another but is always exactly the same in atomistic simulations and in CG simulations, about $40$~kg~m$^{-3}$ for $N=3,5,7,9,23$ and equal to $7.8$~kg~m$^{-3}$ for the case $N=50$.}
  \label{tab:Rgshort}
  \begin{tabular}{c c c c c c}
    \hline
     & atomistic & ImpCG & ExpCG & ExpCG & ExpCG \\
     & & & & -noLJ & -noLJ-noAngle \\
    \hline
    $N=3$ & $2.9$ & $2.3$&$2.3$ &$2.3$ &$2.6$ \\
    $N=5$ & $4.1$& $3.3$& $3.3$& $3.3$&$3.9$\\
    $N=7$ & $4.8$ & $4.1$& $4.1$& $4.3$&$5.0$\\
    $N=9$ & $5.5$ & $4.8$& $4.8$& $5.1$&$6.0$\\
    $N=23$ & $6.9$
    & $8.3$ & $7.4$ & $9.9$ & $11.7$ \\
    $N=50$ & $9.7$
    & $10.8$& $9.7$ & $19.7$ &$ 24$\\
    \hline
  \end{tabular}
\end{table}

The parameters of the ImpCG and ExpCG models were fitted to reproduce the properties of a chain of $23$ monomers at high polymer concentration. To evaluate the transferability of these models to other chain lengths\cite{Carbone08}, we have performed atomistic simulations of PA oligomers made of $3$, $5$, $7$ and $9$ monomers. In each case, the results were averaged over three different atactic chains and the total monomer concentration was about $40$~kg~m$^{-3}$. Also, the properties of one atactic  chain of $50$ monomers at $c=7.8$~kg~m$^{-3}$ was studied from atomistic simulations during $80$~ns.   For all these systems, CG simulations of the salt-free systems at the same concentration were also performed.

The average gyration radius obtained for these systems are collected in Table~\ref{tab:Rgshort}. The gyration radii predicted by the two CG models that are fitted on atomistic results, ImpCG and ExpCG, are exactly the same for $N\le9$, but slightly underestimate the prediction of atomistic simulations in this range. The interest of a CG model is mainly to be able to simulate the behavior of long chains, and it appears that both ImpCG and ExpCG models are able to predict the gyration radius of a chain of $50$ monomers, obtained by atomistic simulations. Nevertheless, it must be noted that, as already observed in Fig. \ref{fgr:histo-Rg-AA}, we observed that the distribution over trajectories of the gyration radius is larger for CG models than for the atomistic description.  

The ExpCG-noLJ and ExpCG-noLJ-noAngle models are not fitted on atomistic simulations, but unravel the effect of two particular characteristics of the ExpCG model, the role of monomer-monomer attractive interactions (1-4 Lennard-Jones), and the role of angular constraints. For chains that are smaller than $23$, both models are remarkably good at predicting the value of the gyration radius. 
The ExpCG-noLJ-noAngle model, which is the simplest one without LJ attraction neither angular constraints, leads to a gyration radius systematically larger than other CG models for every chain length. 
For the chain of $23$~monomers, the predictions of the ExpCG-noLJ and ExpCG-noLJ-noAngle models slightly diverge from the other models, but the divergence becomes intriguingly spectacular for chains of $50$~monomers, thus suggesting an important impact of monomer-monomer attractions on the folding of the chain.  

This last result suggests a different scaling regime for the different CG models. In the last part, we extend our study to longer CG chains, in order to explore in more details the scaling behavior of our CG models.

\subsubsection{Scaling laws predicted by the CG models in the moderate chain length scale }
\label{scaling}

\begin{figure}[ht]
\centering
  \includegraphics[height=7.5cm]{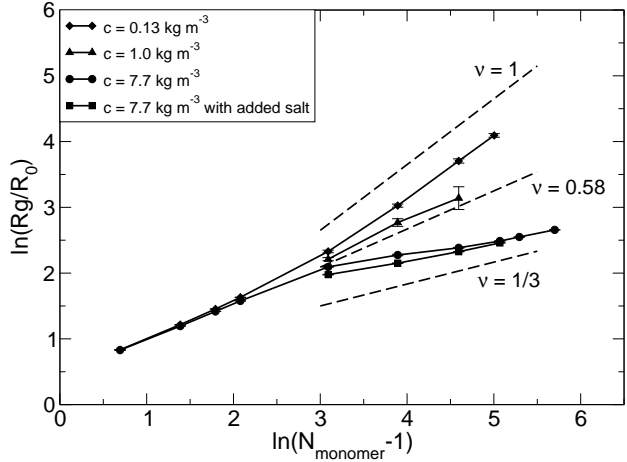}
  \caption{Average gyration radius obtained from the ExpCG model at different monomer concentrations as a function of the number of monomers in a log-log plot.}
  \label{fgr:Rg-vs-N-ExpCG1}
\end{figure}

\begin{figure}[ht]
\centering
  \includegraphics[height=7.5cm]{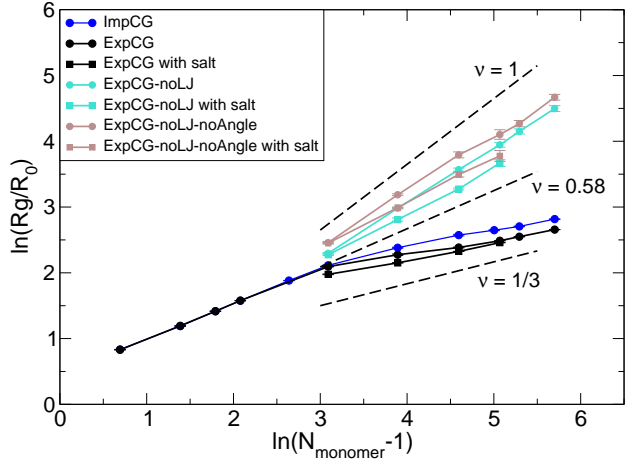}
  \caption{Average gyration radius obtained from different models as a function of the number of monomers in a log-log plot. The concentration of monomers is in any case $c=7.7$~kg~m$^{-3}$.}
  \label{fgr:Rg-vs-N}
\end{figure}

In polymer physics theories, the impact of the type of models, and of the value of the parameters of these models on the polymer properties are primarily studied through the values of the scaling (Flory) exponents. Neutral polymers are well described by the universal class of Interacting Self Avoiding Walks. For such models at equilibrium and in the large-$N$ limit, two different folding modes have been predicted and measured\cite{Nishio79}, depending on the relative strength of the monomer-monomer and solvent-monomer interactions with respect to temperature characterized by the ratio $\varepsilon/k_BT$: In good solvent (low $\varepsilon/k_BT$), the favorable interaction with the solvent leads to an effective repulsion between monomers. Hence, the polymer expands into a decondensed, disordered state called \emph{coil}, described as a \emph{self-avoiding walk} (SAW).  
In poor solvent (high $\varepsilon/k_BT$), monomer-monomer attractions become predominant, and the polymer collapses into a state called \emph{globule}\cite{DeGennes1975}. A phase transition between the two regimes is observed at a specific temperature called  $\Theta$ (\emph{theta}) temperature or $\Theta$ \emph{point}, or equivalently at $\varepsilon_{\theta} = k_B\Theta$.
At this point, the effective repulsion between monomers compensates their attraction\cite{Grosberg1994}. The polymer behaves then as a random walk (RW) (also referred to as $\Theta$-polymer).  
In the case of polyelectrolytes, neutral polymer models can be relevant when the polymer is described at scales which are larger than the Debye length. At lower scales, when electrostatics dominate over thermal fluctuations, monomer-monomer repulsions extend the polyelectrolyte, which takes the shape of a rod.    
The previous folding states correspond to 4 regimes, which are characterized by a specific Flory exponent $\nu$: $1$ for rods (strong electrostatics), $3/5$ for coils (SAW), $1/2$ for theta-polymers (RW), and $1/3$ for globules.

We plot in Fig.~\ref{fgr:Rg-vs-N-ExpCG1} the evolution of the logarithm of the gyration radius as a function of the logarithm of the number of monomers for the ExpCG model at three different polymer concentrations. For all three concentrations, we show the case of salt-free systems, and for the most concentrated system, we also show the influence of an added salt at a fixed concentration equal to $0.14$~mol~dm$^{-3}$.  
As it clearly appears on this plot, the scaling behavior depends on the concentration of the polyelectrolyte. For a salt-free dilute polymer concentration ($c=0.13$~kg~m$^{-3} \simeq  10^{-4}c^\star$), the behavior is close to that of an extended chain ($\nu=1$). It means that at this polymer concentration, for such polymer length, the electrostatic interactions are strong enough to extend the polyelectrolyte chain.
For the intermediate concentration $c=1$~kg~m$^{-3} \simeq 10^{-3}c^\star$, the scaling behavior corresponds to $\nu=0.58$, i.e long range electrostatic repulsions between monomers are screened and the polyelectrolyte behaves as a neutral polymer in the coil regime, where effective interactions only reflect excluded volume effects.
At the highest concentration under study ($c=7.7$~kg~m$^{-3} \simeq 10^{-2}c^\star$), $\nu$ is close to $1/3$, which corresponds to a collapsed polyelectrolyte chain (globule). In the presence of added salt, the behavior is also that of a globule. It indicates the presence of attractive effective interactions between monomers (bad solvent conditions), which overcome electrostatic repulsions. This is not surprising, as we have already shown in the previous parts that for this kind of systems, adding explicit charges to a neutral model with attractive LJ interactions does not lead to an extension of the polymer. 

This spectacular dependence of the scaling regime on the polymer concentration is typical of salt free systems. The counterion concentration is equal to the monomer concentration, and therefore the Debye length scales as the inverse square root of the polymer concentration. 
If one looks more carefully at the curves, there are two signs of breaking of the aforementioned scaling laws, which are due to finite size effects. The first indication is the change of slope of $\ln(Rg)$ as a function of $\ln(N-1)$ when $N$ is above $23$ ($\ln(N-1)=3$). The second indication is the existence of domains where the slope seems lower than $1/3$, which does not exist in the thermodynamic limit $N \to \infty$. Such observations are similar to what is known for finite size neutral polymers undergoing a coil globule transition. At a fixed interaction potential, and at a fixed temperature below the theta point, small polymers can behave as coils, while longer polymers behave as globules. Therefore, the scaling exponent decreases as $N$ increases. Moreover, there exists a region of crossover between coil and globule where the effective scaling exponent can be smaller than $1/3$\cite{Grassberger1995,Care2014}.
The behavior of the ExpCG model in the high polymer concentration regime is consistent with finite size polymers at the coil-globule crossover: The shortest chains behave as coils with a $0.58$ exponent, then for $N$ greater than 23, the scaling law breaks and there is a crossover between coils and globule characterized by an effective scaling exponent lower that $1/3$. 

The implicit ion coarse grained model, ImpCG, behaves similarly to the concentrated ExpCG model, as it can be seen in Fig.~\ref{fgr:Rg-vs-N}. In other words, both the ImpCG and the ExpCG models at a concentration  $c\simeq 10^{-2}~c^\star$ belong to the universal class of Interacting Self Avoiding Walks close to the coil globule transition. Indeed, both models include a 1-4 LJ attraction. Also, the Debye screening length is short at this concentration, so that the electrostatic repulsion between monomers is screened. Nevertheless, in the case of the ImpCG model, the deviation from the SAW scaling law ($\nu = 0.58$) is slightly less important than with the explicit ions model, which correlates with the greater extension of long chains ($N > 50$) with this model. It should be reminded that the molecular dynamics simulations used to calibrate the different coarse-grained models are even more concentrated ($c=39$~kg~m$^{-3}$).

More remarkably, the ImpCG model is not able to predict the scaling behavior of polyelectrolytes at the two lowest polymer concentrations. Indeed, according to explicit charge models, the increase of electrostatic interactions at low polymer concentrations totally changes the phase behavior of the chain, which is no longer close to the coil-globule transition. We clearly see here the difficulty of using neutral models (such as the ImpCG model) for salt-free polyelectrolytes: (1) If the potential is fitted in a concentration domain corresponding to a given scaling regime, there is no chance the same model can describe the system in a different scaling regime; (2) due to the size limitations of systems described by molecular dynamics, implicit CG models are necessarily restricted to concentrated domains. It is interesting to note that the addition of screened electrostatic repulsions at the Debye-H\"uckel level to ImpCG also fails to predict the coil regime at the intermediate concentration: It predicts a rod regime contrarily to the more accurate explicit charge model.  

More generally, the previous results suggest that, even in the presence of salt, the only domains for which implicit ion models are relevant are those for which the polyelectrolytes behave as a neutral polymer. Indeed, if long range electrostatics are important, (1) there is no chance an implicit ion model fitted with molecular dynamics, with specific values of the polymer and salt concentrations, is transferable to other concentrations, (2) the CG model will be restricted to very concentrated systems. 

The results obtained with the explicit ion model but without LJ interactions or without angular interactions, ExpCG-noLJ and ExpCG-noLJ-noAngle, are displayed on Fig. \ref{fgr:Rg-vs-N}. Among with implicit ion models, the polymer modeling community often resorts to simplified models, such as our ExpCG-noLJ model without Lennard Jones attractions, and the ExpCG-noLJ-noAngle without angular constraints. These models, that do not include any direct attraction between distant beads, at the highest concentration $c\simeq 10^{-2}~c^\star$, display a scaling behavior between coils ($\nu=0.58$) and rods ($\nu=1$), even in the presence of added salt. 
These models are not able to reproduce the coil-globule transition seen with the ExpCG and ImpCG models. This is not a surprise, but it demonstrates that the screening effects of counterions or of an added salt at $0.14$~mol~dm$^{-3}$ are not sufficient to yield the folding of the chain into a polymer that is more compact than a random coil (SAW). In other words, in these systems, like charge attractions are not present for these CG models.

\section{Conclusions}

We have designed several coarse-grained models of the sodium polyacrylate based on atomistic simulations of this system. The explicit charge model, which better reproduces the atomistic properties of a short chain ($N=23$) seems good at predicting the mean gyration radius of a longer polymer ($N=50$), as well as shorter chains. To account for the radial distribution functions obtained at the atomistic level, we had to include an attractive Lennard-Jones contribution to the interaction between distant monomers, even in the case where counterions are explicitely described. It means that the electrostatic screening of counterions in salt-free systems is not sufficient to reproduce the effective attraction between monomers. Usual simple models of polyelectrolytes do not include such 1-4 LJ attractions. We have also shown that implicit ion models are relevant only to  concentrated solutions of polymers or to the case where electrostatics is screened by an added salt.  

Building coarse-grained models of polyelectrolytes from atomistic simulations is a challenging task for several reasons.
First, atomistic simulations of polyelectrolytes are restricted to relatively short polymers, as they combine the difficulties of polymers (long relaxation times) and electrolytes (long-range interactions). As we have seen in the present study, this limitation imposes the determination of the mesoscopic model in a polymer size domain in which the finite-size effects are important. Moreover, the size of atomistic simulation boxes is limited to relatively short values, where box size effects are important. Second, the behavior of the atomistic model is very sensitive to small changes either in the force-field\cite{MintisJPCB2019}, or in the stereoisomery of the monomers, and very long trajectories are needed to converge to equilibrium states as we have seen in the present study. Nevertheless, once we have a reliable coarse-grained model of a given polyelectrolyte in implicit water, the gain in computation time is substantial and allows one to investigate the dynamic properties of the chain at larger length and time scales. Our goal in future works is to use mesoscopic simulation method including hydrodynamic couplings to compute the dynamic properties of these systems in bulk and under a confinement.

\section{Acknowledgements}
The authors acknowledge financial support from the French National
Agency for Research (ANR) under grant ANR-09-JCJC-0082-01, from Sorbonne Universit\'e, and from the CNRS (french National Centre for Scientific Research).

\providecommand{\latin}[1]{#1}
\makeatletter
\providecommand{\doi}
  {\begingroup\let\do\@makeother\dospecials
  \catcode`\{=1 \catcode`\}=2 \doi@aux}
\providecommand{\doi@aux}[1]{\endgroup\texttt{#1}}
\makeatother
\providecommand*\mcitethebibliography{\thebibliography}
\csname @ifundefined\endcsname{endmcitethebibliography}
  {\let\endmcitethebibliography\endthebibliography}{}

\end{document}